# Elemental (im-)miscibility determines phase formation of multinary nanoparticles co-sputtered in ionic liquids

Michael Meischein,[a] Alba Garzón-Manjón,[b] Thomas Hammerschmidt,[c] Bin Xiao,[a] Siyuan Zhang,[b] Lamya Abdellaoui,[b] Christina Scheu,[b] and Alfred Ludwig*[a]

[a] *Chair for Materials Discovery and Interfaces, Institute for Materials, Faculty of Mechanical Engineering, Ruhr University Bochum, Universitätsstr. 150, D-44780 Bochum, Germany.*
[b] *Max-Planck-Institut für Eisenforschung GmbH, Max-Planck-Straße 1, D-0237 Düsseldorf, Germany.*
[c] *Chair of Atomistic Modelling and Simulation, Interdisciplinary Centre for Advanced Materials Simulation (ICAMS), Ruhr University Bochum, Universitätsstr. 150, D-44780 Bochum, Germany.*

**Abstract**

Non-equilibrium synthesis methods allow to alloy bulk-immiscible elements into multinary nanoparticles, which broadens the design space for new materials. Whereas sputtering onto solid substrates can combine immiscible elements into thin film solid solutions, this is not clear for sputtering of nanoparticles in ionic liquids. Thus, the suitability of sputtering in ionic liquids for producing nanoparticles of immiscible elements is investigated by co-sputtering the systems Au-Cu (miscible), Au-Ru and Cu-Ru (both immiscible), and Au-Cu-Ru on the surface of the ionic liquid 1-butyl-3-methylimidazolium bis-trifluoromethylsulfonyl)imide [Bmim][(Tf)$_2$N]. The sputtered nanoparticles were analyzed to obtain (i) knowledge concerning the general formation process of nanoparticles when sputtering onto ionic liquid surfaces and (ii) information, if alloy nanoparticles of immiscible elements can be synthesized as well as (iii) evidence if the Hume-Rothery rules for solid solubility are valid for sputtered nanoparticles. Accompanying atomistic simulations using density-functional theory for clusters of different size and ordering confirm that the miscibility of Au-Cu and the immiscibility of Au-Ru and Cu-Ru govern the thermodynamic stability of the nanoparticles. Based on the matching experimental and theoretical results for the NP/IL-systems concerning NP stability, a formation model of multinary NPs in ILs was developed.

## 1 Introduction

Metal nanoparticles (NPs) are useful for applications, e.g. in medicine,[1–4] optics,[5] electronics[6] and catalysis[7,8]. Particularly important for the properties of the NPs are their composition and the corresponding elemental distribution within the NPs, their crystal structure and their morphology.[9] Multinary nanoparticles are of high interest, as they can provide excellent functional properties, e.g. as high entropy catalysts.[10,11] The search space for multinary NPs is already large and open questions are related to combinations of elements which are not miscible in the bulk. In nanoscale solid solutions, synthesized by non-equilibrium methods, such combinations of bulk-immiscible elements can provide functionalities which would be otherwise not accessible.[12,13] E.g., new and effective electrocatalysts can be identified by screening a large range of homogeneously alloyed NPs of different compositions but identical crystal structure and nearly identical nanostructure, also partially composed of immiscible elements.[12]

Synthesis of NPs is achievable by a large variety of processes which can be categorized in thermodynamic equilibrium processes and non-equilibrium processes. Equilibrium synthesis routes like thermal decomposition, seed-mediated growth, co-reduction or galvanic replacements lead for immiscible systems with positive heats of mixing to multi-phase structures, e.g. core-shell NPs.[12] Non-equilibrium processes (e.g. carbothermal shock synthesis[11]) can kinetically trap thermodynamically non-miscible elements in a "forced solid solution". E.g. if the necessary energy for sufficient atom mobility is only provided for a short time, an ordering of the immiscible elements according to the thermodynamic equilibrium is not possible, as shown by Yang et al.[12]

Sputtering onto solid substrates at low temperatures is a non-equilibrium process and can be used for synthesizing forced solid solutions. A targeted multinary composition can be synthesized by adjusting the sputter rates of the individual (elemental) targets in a co-sputter process.[14,15] Even elements which are thermodynamically not miscible can be combined in the form of multinary forced solid solutions.[14] Moreover, sputtering onto the surface of ionic liquids (ILs) is a versatile method for synthesizing multinary NPs. However, it is not clear if the non-equilibrium co-sputter process also yields multinary NPs composed of immiscible elements when sputtering on IL. While the co-sputter process controls the amount and ratios of elements arriving at the liquid substrate, composition, crystal structure, size and morphology of the NPs is influenced by the used ILs.[16–20]

ILs are salts which are liquid at temperatures < 100°C, composed of cations and anions with molecular structures enabling the liquid state at ambient conditions. From theoretically 10$^{18}$ different ILs[21] optimal task-specific ILs can be selected for an application. ILs have outstanding possibilities as solvent and synthesis and reaction media for NPs. An important characteristic of ILs is their negligible vapor pressure[22,23] which enables their application as clean and pure substrates for ultrahigh vacuum processes[24]. These special chemical and physical characteristics make ILs interesting for applications in industry[25,26] as reaction and scavenging media and for catalysis in general[27,28]. Additionally, their consideration as green solvents[29] demonstrates their potential for decreasing the environmental stress caused by those industrial applications if the conventional solvents are replaced by ILs.

In this paper, we want to clarify the possibilities of sputtering onto ILs for the formation of multinary solid solution NPs comprising (im-)miscible elements. Co-sputter deposition from elemental targets, simultaneously onto IL and solid substrate reference samples were conducted. The exemplary system Au-Cu-Ru and its subsystems were investigated as the bulk phase diagrams of the binary subsystems show relevant differences: Au-Cu shows complete solid solubility for temperatures > 410°C.[30] At lower temperatures, the formation of the intermetallic phases Au$_3$Cu, AuCu (tetragonal or orthorhombic) and AuCu$_3$ can occur.[30] The system Cu-Ru shows no solubility below 1059°C. The Au-Ru system shows very limited solubility (< 5 at.% Ru in Au) at temperatures between 900°C and 950°C,[31] otherwise the system is immiscible. No ternary phases of the system Au-Cu-Ru are known.

These differences can be rationalized by looking on the characteristics of the used elements, listed in Table 1. According to the Hume-Rothery rules[32,33] two different elements likely form a solid solution when the crystal structures match, the atomic radii of the involved elements do not differ by more than 15%, the difference in electronegativity is small to avoid the formation of compounds and the valence of both elements is similar. For the investigated Au-Cu-Ru system, the differences in atomic radius is < 15%, the difference in electronegativity is < 0.7 and each element has a common valence with the other two elements. However, the crystal structure of Ru (hcp) differs from Au and Cu (both fcc), i.e. one demand of the Hume-Rothery rules is violated.

All binary subsystems have been prepared in the nanoscale as solid solution NPs, although the immiscible Au-Ru and Cu-Ru systems should decompose according to thermodynamics. Those solid solution NPs were applied in catalysis where they could outperform the single elements in specific reactions. E.g. binary Au-Cu NPs are more effective in CO oxidation[38] and propene epoxidation[39,40] than pure Au and Cu NPs, whereas Cu-Ru NPs excel single Cu and Ru NPs as hydrogenation catalyst[41] and pure Ru NPs for CO oxidation[42]. An electrode of a methanol oxidation fuel cell modified with Au-Ru NPs showed superior efficiency when compared to an electrode modified with Ru NPs.[43] Selective control of fcc and hcp crystal structures in Au–Ru solid-solution alloy NPs ($AuRu_3$, fcc and hcp) was achieved by a polyol reduction process.[44] Thus, the better activity of the binary NPs from the Au-Cu-Ru system confirm the hypothesis of a better performance of NPs with increasing compositional complexity,[10] but the stability of the bulk-immiscible systems Au-Ru and Cu-Ru contradicts the principles of miscibility according to the Hume-Rothery rules.

However, in the nanoscale other principles and effects can dominate the material characteristics compared to bulk materials:[45–48] The stability of nanoscale material systems can depend on different aspects of the selected synthesis process or the (multinary) system itself. The formation of stable NPs is determined by their thermodynamic state, and is influenced by their size, shape and crystal structure and for multinary NPs also by their composition.[49] For sputtering multinary thin films, even metastable alloys of immiscible elements can be realized and their stability against decomposition can be studied e.g. by the combinatorial processing platform approach.[50] Given these considerations, the composition and structure of the co-sputtered NPs stabilized in ILs will reveal if sputtering performed as a non-equilibrium process onto ILs is applicable for obtaining multinary NP of immiscible elements.

The synthesized Au-Cu-Ru materials (films and NPs) were investigated by energy-dispersive X-ray spectroscopy (EDS) and X-ray diffraction (XRD) analysis of the thin films which formed on wafer pieces added to the depositions on the ILs as well as inductively coupled plasma mass spectrometry (ICP-MS) of the obtained NP/IL-suspensions and transmission electron microscopy (TEM) and EDS analysis of the NPs. The theoretical probability of the formation of multinary NPs comprising the elements from the Au-Cu-Ru system was evaluated in terms of the thermodynamic stability as obtained by atomistic simulations with density functional theory (DFT). Besides the information concerning elemental miscibility influence on NP stability and the effect of the synthesis process (equilibrium vs. non-equilibrium), the comparison between elemental NP, thin film and IL composition will reveal further information concerning the NP formation process when sputtering onto ILs[51]. This information is important for design and synthesis of next-generation nanoscale materials.

## 2 Experimental Methods

### 2.1 Synthesis of binary and ternary thin films and NP/IL-suspensions by co-sputtering

The IL 1-butyl-3-methylimidazolium bis(trifluoromethylsulfonyl)imide [Bmim][(Tf)$_2$N] (Iolitec, Heilbronn, Germany) was used for all sputter depositions. The IL purity was > 99% (halides content < 100 ppm, water content 51 ppm). No further processing of the IL was performed prior to sputter depositions. IL was stored and handled under Ar atmosphere in a glovebox (oxygen and water content both < 0.5 ppm). A commercial co-sputter system (AJA POLARIS-5, *AJA INTERNATIONAL, Inc.*, North Scituate, MA, USA) with 1.5-inch diameter magnetron sputter cathodes and multiple sputter source DC power supplies (DC-XS 1500 and DC-XS 750 from *AJA INTERNATIONAL, Inc.*, North Scituate, MA, USA) was used. Elemental targets (Au, Sindlhauser Materials, Kempten, Germany; Cu and Ru both EvoChem, Offenbach am Main, Germany), all with purity 99.99% and dimensions 38.1 mm diameter × 4.775 mm thickness were applied. For sputtering, Ar (Praxair, Düsseldorf, Germany) with purity 99.9999% was used.

For sputter deposition, the IL was placed in a custom-made cavity plate with adaptable lids for covering unfilled cavities (see Meyer et al.[52]) to expose the IL surface to the flux of sputtered atoms. Prior to the deposition, plate and lid were cleaned in an ultrasonic bath for 30 min in technical acetone (purity ≥ 99.5%) and isopropanol (purity ≥ 99.7%) respectively and dried in an oven at 80°C for 1 h. Each exposed cavity was filled with a volume of 35 μL IL inside the glovebox for each deposition. A piece of patterned Si/SiO$_2$ wafer (photolitographically structured with a photoresist lift-off cross pattern for thin film thickness determination) was placed on the lid next to the cavities for each deposition onto ILs to measure the composition and thickness of the films. Thin film compositions were analyzed by EDS (Oxford Instrument X-act) in a scanning electron microscope (SEM, JEOL JSM-5800 LV). The crystal structure of the thin films was investigated by XRD using a Bruker D8 Discover diffractometer with an IμS microfocus source (50 W, Cu K$_\alpha$ radiation at $\lambda \cong 0.15406$ nm) and a VÅNTEC-500 2D detector. To cover the 2θ range of interest, three frames were captured in the offset couple 2θ/θ scan type. The start angles for the three frames were θ = 12°/2θ = 30° with stop angles θ = 42°/2θ = 90° and increment θ = 15°/2θ = 30° and an integration time of 60 s for each frame (sample to detector distance of 174.6 mm, collimator diameter of 1 mm). Literature data (except for Au-Ru) for comparison were obtained from the ICSD database. The data sets were used for simulating XRD diffraction patterns for Cu K$_\alpha$ radiation ($\lambda \cong 0.15406$ nm) using the VESTA software for visualizing crystallographic data (see K. Momma and F. Izumi[53]).

**Table 1:** Elemental data of Au, Cu and Ru, taken from Mizutani[32] and Hume-Rothery et al.[33] if not noted differently by the addition of literature sources at the specific column heading.

| Element | Crystal structure | Atomic radius (pm)[34] | Electronegativity (Pauling scale)[35] | Valence[34] | Lattice parameters / reduced cell (Å)[36,37] | Difference of lattice parameters (%) | Weighted surface energy (J/m$^2$)[36,37] | Molar heat of vaporization (kJ/mol)[34] |
|---|---|---|---|---|---|---|---|---|
| **Au** | fcc | 134 | 2.54 | 1+, 3+ | a = 2.950<br>b = 2.950<br>c = 2.950 | to Cu for a = b = c: 15.2<br>to Ru for a = b: 7.9<br>to Ru for c: 31.6 | 0.75 | 334.4 |
| **Cu** | fcc | 117 | 1.90 | 1+, 2+ | a = 2.561<br>b = 2.561<br>c = 2.561 | to Au for a = b = c: 13.2<br>to Ru for a = b: 6.3<br>to Ru for c: 40.6 | 1.42 | 300.7 |
| **Ru** | hcp | 125 | 2.20 | 2+, 3+, 4+, 6+, 8+ | a = 2.733<br>b = 2.733<br>c = 4.314 | to Au for a = b: 7.4<br>to Au for c: 46.2<br>to Cu for a = b: 6.7<br>to Cu for c: 68.4 | 2.88 | 595.5 |

Prior to all depositions, the IL in the cavity plate was evacuated for at least 72 h for removing remaining oxygen and water from the transport out of the glovebox into the sputter chamber resulting in an exposure of the IL towards air of about 10 s. Subsequent to plasma ignition (parameters in Table 2), the targets were pre-cleaned with closed shutters for 120 s, a rotation of the cavity plate of 30 rotations per minute (for obtaining a homogenous composition over the whole area) and a successive reduction of the Ar pressure to the deposition pressure. Pre-cleaning of the targets was performed for removing possible oxide or organic contamination layers[52] followed by adjusting the deposition power (see Table 2) and opening the shutters in front of the targets for the specific deposition times. A tilt of the cathodes resulted in an angle of 12° between the target normal and the normal of the cavity plate. For all depositions, a film thickness of 500 nm and the composition denoted in Table 2 for the sputtered thin film on the wafer piece was adjusted. After each deposition, the cavity plate was transferred immediately into the glovebox for collection, storage and processing of the sputtered NP/IL-suspensions.

## 2.2 TEM sample preparation

The sputtered NP/IL-suspensions were prepared for TEM investigations by dropping 2.5 μL of each IL on the carbon-coated side of holey carbon-coated Au grids (200 mesh, Plano GmbH, Wetzlar, Germany). The ILs were left at this side for 2.5 h to enable adhesion of the NPs on the carbon film. Washing the grids dropwise with dried acetonitrile for 1 h under inert conditions (see Meyer et al.[52]) was conducted to prevent grid contamination originating from the IL being exposed to the electron beam during TEM investigations. The washed grids were stored in the glovebox until the specific TEM investigation. At least 198 NPs per NP/IL-suspension were counted for evaluating the NP diameters. Conventional TEM and high-resolution TEM (HRTEM) studies were performed with a Tecnai F20 S/TEM instrument (Thermo Fischer Scientific, Eindhoven, Netherlands) operated at 200 kV. High angle annular dark field (HAADF) STEM imaging and STEM-EDS spectrum imaging were performed on an aberration-corrected TEM (Titan Themis) equipped with a Bruker Super X detector (Thermo Fischer Scientific, Eindhoven, Netherlands), operated at 300 and 120 kV. Prior to the EDS and HAADF investigations the TEM grids were cleaned with a "Femto"-series plasma cleaner (Diener electronic GmbH & Co. KG, Ebhausen, Germany) in a plasma composed of 75% Ar and 25% O$_2$ for 30 s at 50 W immediately before inserting them in the instrument to remove hydrocarbon contaminants. The HAADF detector covered a range from 73 to 200 mrad for 300 kV and from 95 to 200 mrad for 120 kV.

EDS measurements were performed with currents of ~150 pA and using an electron probe with a beam size of ~0.1 nm and a convergence semiangle of 23.8 mrad. Multivariate statistical analysis was performed on the STEM-EDS data to reduce the noise and facilitate the detection of Ru.[54]

## 2.3 ICP-MS measurements

For the ICP-MS measurements, an amount 50 μL of each IL was separated in suitable Teflon containers. The separated samples were diluted each with 4 ml of 69% concentrated phosphoric acid of the "ROTIPURAN® Supra" line. The obtained mixtures were chemically digested in a Multiwave Pro microwave digestion device with 8-slot container holder 8NXF100 (from Anton Paar GmbH, Graz, Austria). The digestion occurred at maximum temperature 240°C and maximum pressure 6 MPa to guarantee a complete transfer of the investigated material into solvable nitrides which could be measured in the ICP-MS. Subsequent to the digestion, the resulting solutions were further diluted with ultrapure water (conductivity 0.055 μS/cm) to a total volume of 10 ml per sample. From this stem-solutions, 1:100- and 1:1000-dissolutions were produced and experienced an acidification with 2% phosphoric acid prior to the measurements in an iCAP RQ ICP-MS device (from Thermo Fisher Scientific Inc., Waltham, Massachusetts, USA). The ICP-MS measurements were performed in the KED-mode to decrease disturbances from molecule ions.

Table 2: Sputter deposition parameters for depositions onto ILs and reference wafer pieces.

| Deposition | Start pressure (Pa) | Ignition pressure (Pa) | Ignition power (W) | Pre-clean duration (s) | Deposition pressure (Pa) | Deposition power (W) | Deposition duration (min) | Targeted composition (at. %) |
|---|---|---|---|---|---|---|---|---|
| Cu-Ru | $1.12 \times 10^{-4}$ | 1.33 | 20 | 120 | 0.5 | 30 (Cu) 35 (Ru) | 50 | 50:50 |
| Au-Ru | $1.03 \times 10^{-4}$ | 1.33 | 20 | 120 | 0.5 | 9 (Au) 30 (Ru) | 70 | 50:50 |
| Au-Cu | $1.16 \times 10^{-4}$ | 1.33 | 20 | 120 | 0.5 | 15 (Au) 30 (Cu) | 45 | 50:50 |
| Au-Cu-Ru | $1.16 \times 10^{-4}$ | 1.33 | 20 | 120 | 0.5 | 9 (Au) 17 (Cu) 30 (Ru) | 52 | 33:33:33 |

## 2.4 DFT calculations

DFT calculations of bulk systems and nanoparticles were carried out with the VASP software[55–57] by a high-throughput framework[58] using the projector-augmented-wave method[59] and the Perdew-Burke-Ernzerhof exchange-correlation functional[60]. All calculations were non-spinpolarized and used a cutoff of 400 eV. The DFT calculations of bulk systems include Monkhorst-Pack k-point meshes[61] with a distance of 0.02 Å$^{-1}$. With these settings, very good agreement with previous DFT calculations[62] regarding, e.g., the equilibrium lattice constant was achieved. The (im-)miscibility of the binary subsystems of Au-Cu-Ru was investigated with typical ordered structures based on bcc (PdTi, AlOs, VZn, B2, B11, C11$_b$), fcc (Al$_8$Ni, D0$_{22}$, D0$_{23}$, L1$_0$, L1$_2$, L6$_0$) and hcp (B8$_1$, B8$_2$, B35, B$_b$, B$_h$, B$_i$) lattices. The possibility of solid-solution phases was taken into account with special-quasi-random structures (SQS)[63]. Using the ATAT package[64], five 2×2×2 fcc unit cells with 32 atoms were generated for the compositions AB, AB$_3$, A$_3$B and A$_6$BC and the numerical average of the corresponding DFT results for the five SQS structures was taken.

The DFT calculations of the NPs used Γ-centered k-point meshes and relaxation of the atomic positions. The NP structures were generated by spherical cutouts from periodic repetitions of fcc unit cells. These shapes correspond to the experimentally observed spherical and crystalline NPs presented in the following. The resulting NP structures contain 16, 28, 44, 68, 80, 104 and 140 atoms with radii that depend on the content of Au, Cu and Ru atoms. The relative initial radii are 1.0, 1.1, 1.3, 1.5, 1.7, 1.8 and 2.0 in terms of the fcc lattice constant of the particular chemical composition. For the 140-atom AuCu NP with L1$_0$ ordering, e.g., we obtain a radius of 1.4 nm, comparable to the experimentally observed NPs. The unit cells of the DFT calculations for the NP are chosen to include a separation of approximately 1 nm to the periodic images of the NP which leads to simulation cells of 1.6 nm × 1.6 nm × 1.6 nm for NP with 16 and 28 atoms, 2 nm × 2 nm × 2 nm for NP with 44, 48 and 80 atoms, and 2.4 nm × 2.4 nm × 2.4 nm for NP with 104 and 140 atoms.

# 3 Results

Experimental results of the thin film and NP analysis were compared with results from DFT simulations. The theoretical consideration of the NP stability helps to understand and also validates the observed NP compositions and the NP formation.

## 3.1 Analysis of the thin films

The co-sputtered thin films on solid substrates were used to control if the desired equiatomic compositions were reached. The film compositions are listed in Table 3 and show a satisfying accuracy to the aimed compositions in Table 2 since the biggest difference from equiatomic compositions was 2.8 at.% for the binary and 2.2 at.% for the ternary systems. The film compositions serve as reference for comparison with the compositions of the NP/IL-suspensions.

**Table 3:** Compositions of co-sputtered thin films. The values represent the average composition of ten EDS measurement points arranged on a straight line.

| System | Composition (at.%) | | |
|---|---|---|---|
| | Au | Cu | Ru |
| Au-Cu | 47.2 | 52.8 | - |
| Au-Ru | 48.5 | - | 51.5 |
| Cu-Ru | - | 53.4 | 46.6 |
| Au-Cu-Ru | 35.1 | 33.8 | 31.1 |

The capability of sputtering to force bulk-immiscible elements into thin film solid solutions (i.e. deposition on solid substrates) is investigated by the XRD analysis of the thin films which were simultaneously obtained with the co-depositions onto ILs. The results are illustrated in Figure 1. According to the XRD patterns, all four thin films show fcc solid solutions in agreement with literature. The bump of the peak for the (111) lattice plane in Figure 1 (a) can be attributed to a low signal-to-noise ratio since a long exposure time for this system had to be used for obtaining reasonable counts.

## 3.2 Analysis of nanoparticles stabilized in ILs

Co-sputtering onto an IL surface results in an accumulation of the sputtered atoms in the IL accompanied with the formation of NPs. In a previous study on co-sputtering Au and Cu in IL[65], thin films on wafer pieces (added to the depositions) were used to determine the sputter flux which reaches the IL surface (thin film composition Au$_{53}$Cu$_{47}$). However, the NPs showed compositions ranging from Au$_{75}$Cu$_{25}$ to Au$_{87}$Cu$_{13}$), i.e. a clear compositional difference of thin film and NPs was observed. These differences contradicted the primary idea of this previous study that the NP composition must accord with the composition of the vapor sputtered onto the solid substrates and onto the IL. To unravel the origin of the deviant NP composition with respect to the sputtered thin films, an analysis of the IL composition in terms of metal contents is necessary. To this purpose, ICP-MS measurements of the co-sputtered ILs were conducted, which could be compared with the NP compositions. The ICP-MS results are listed in Table 4. The compositions of the co-sputtered NP/IL-suspensions are in good agreement with the thin film compositions (Table 3). Thus, the sputtered atoms do penetrate in the IL volume and are not reflected at the IL surface or do not show a different diffusion behavior for the individual elements from the sputter vapor phase into the ILs, as assumed in the previous study.

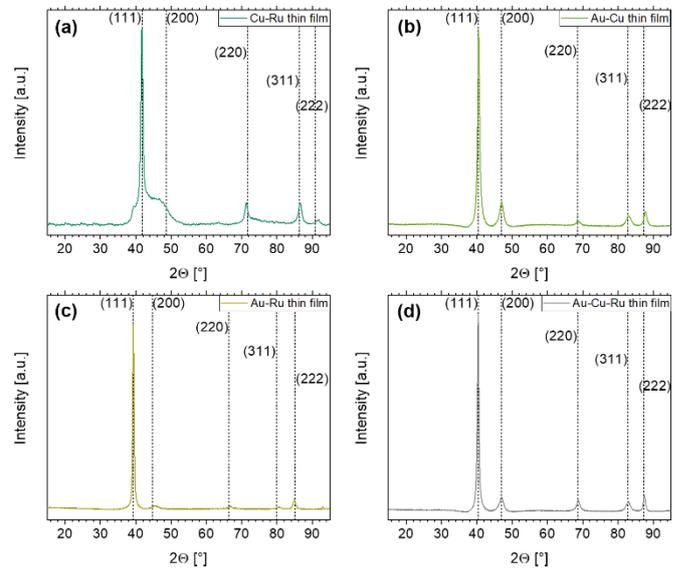

**Figure 1:** XRD patterns of the co-sputtered thin films. Dashed lines indicate literature data for the systems with the corresponding lattice planes annotated in round brackets. For the Cu$_{53.4}$Ru$_{46.6}$ thin film in (a), the reference data were obtained from fcc Cu-Ru with equiatomic composition (space group Fm-3m).[66] For the Au$_{47.2}$Cu$_{52.8}$ thin film in (b), the reference data were obtained from powder diffraction experiments with equiatomic fcc AuCu powder (space group Fm-3m).[67] The reference data for the Au$_{48.5}$Ru$_{51.5}$ thin film in (c) were obtained from Au$_{47.5}$Ru$_{52.5}$[68] analyzed with a conventional XRD system (Cu K$_\alpha$ radiation). For the Au$_{35.1}$Cu$_{33.8}$Ru$_{31.1}$ thin film in (d), the same reference as for the Au-Cu thin film was used.

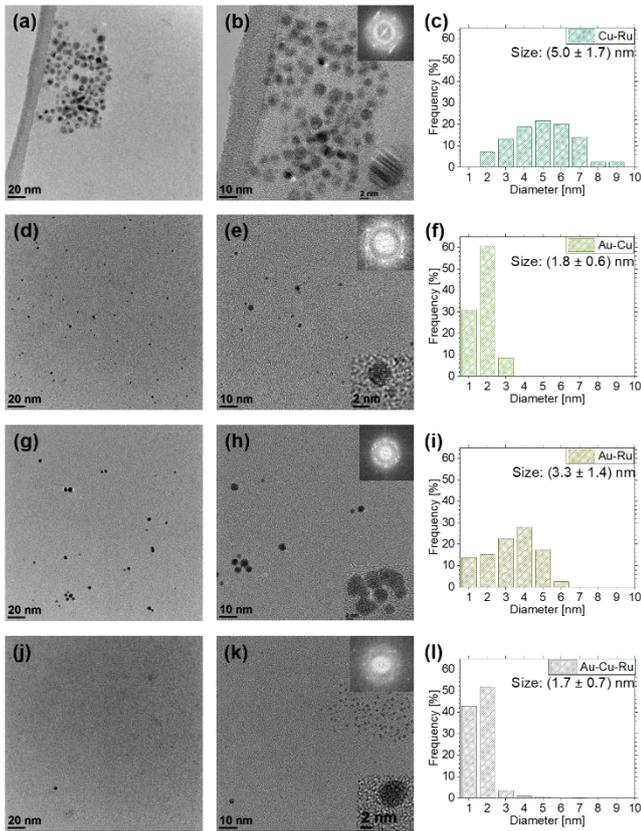

**Figure 2:** TEM images of co-sputtered NP for the systems Cu-Ru in (a) and (b), Au-Cu in (d) and (e), Au-Ru in (g) and (h) and Au-Cu-Ru in (j) and (k). Crystallinity of the NPs was confirmed by FFT of the corresponding HRTEM images depicted in the insets. The corresponding size distributions with mean diameter for each binary and the ternary co-sputtered NPs are depicted next to the corresponding TEM images in (c) for Cu-Ru, in (f) for Au-Cu, in (i) for Au-Ru and in (l) for Au-Cu-Ru.

**Table 4:** Compositions of the co-sputtered NP/IL-suspensions as analyzed by ICP-MS. The compositions of the co-sputtered thin films were: $Au_{47.2}Cu_{52.8}$, $Au_{48.5}Ru_{51.5}$, $Cu_{53.4}Ru_{46.6}$ and $Au_{35.1}Cu_{33.8}Ru_{31.1}$.

|  |  | Co-sputtered systems | | | |
|---|---|---|---|---|---|
|  |  | Au-Cu | Au-Ru | Cu-Ru | Au-Cu-Ru |
| **Elemental composition (at.%)** | Au | 49.1 | 46.8 | - | 36.8 |
|  | Cu | 50.9 | - | 52.6 | 33.0 |
|  | Ru | - | 53.2 | 47.4 | 30.2 |

TEM investigations were performed to correlate the NP composition, the composition of the sputtered IL and the NP size and shape. The obtained TEM data and previous investigations of the authors show that co-sputtering from elemental targets on IL can lead to the formation of multinary NPs.[14,65] However, composition, size and size distributions differ for the binary and ternary NP systems, see Figure 2 and 4.

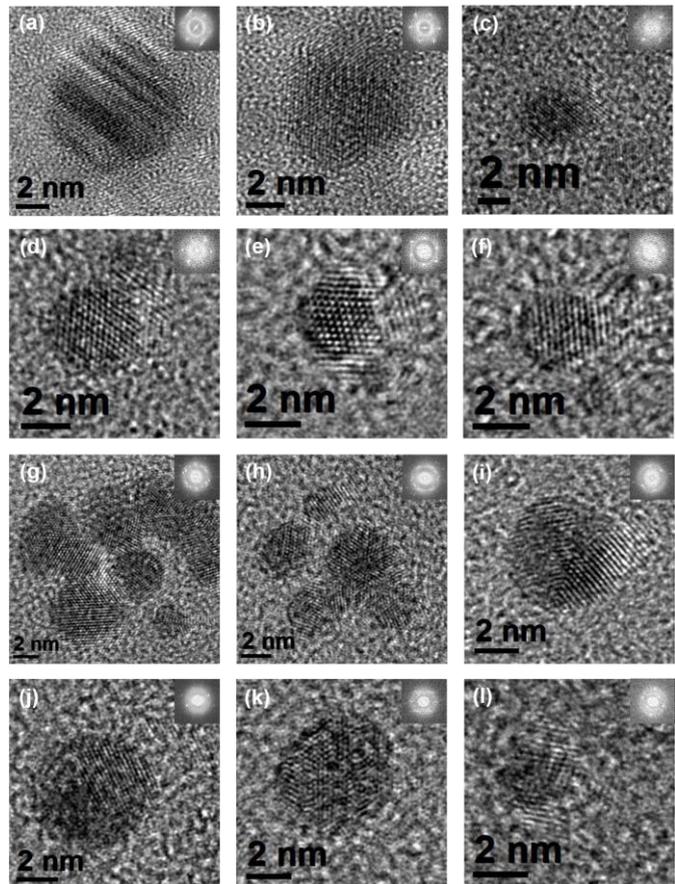

**Figure 3:** HRTEM images of co-sputtered NPs. (a), (b) and (c): examples of Cu-Ru NPs; (d), (e) and (f): examples of Au-Cu; (g), (h) and (i): examples of Au-Ru; (j), (k) and (l): examples of Au-Cu-Ru. Crystallinity of the NPs is visible as lattice fringes in the HRTEM images and was confirmed by FFT (insets). The NPs show different orientations of the crystal structures as representation of the different miscibility of the constituent elements.

For Cu-Ru and Au-Ru, which are immiscible in bulk, the NP size distributions are much broader than for Au-Cu and Au-Cu-Ru NPs and their mean size (Cu-Ru: $(5.0 \pm 1.7)$ nm, Au-Ru: $(3.3 \pm 1.4)$ nm) is nearly twice the mean size of the miscible binary Au-Cu and the NPs from the ternary IL (Au-Cu: $(1.8 \pm 0.6)$ nm, Au-Cu-Ru: $(1.7 \pm 0.7)$ nm).

For all four co-sputtered systems the NPs appear crystalline and spherical (see Figure 3). Lattice fringes aligned in different directions indicate the existence of areas of different crystallographic orientations within individual NPs.

Figure 4 shows the compositions of the co-sputtered NPs from each binary system. The denoted compositions represent the signals of a round mask covering the complete NP without background. The amount of variation of the set values was calculated using a low standard deviation ($1\sigma$, which is close to the expected value). Concerning the elemental miscibility, clear differences are visible. For the bulk-immiscible combinations Cu-Ru in Figure 4 (a) and (b) and Au-Ru in Figure 4 (e) and (f), compositions of $Cu_{94}Ru_6$ to $Cu_{100}Ru_0$ for the Cu-Ru IL and $Au_{96}Ru_4$ to $Au_{99}Ru_1$ for the Au-Ru IL were identified. Ru atoms appear to be only present at the surfaces of the NPs without building a closed shell. For the combinations of bulk-miscible elements, Au-Cu in Figure 4 (c) and (d) and Au-Cu-Ru in Figure 4 (g) and (h), compositions of $Au_{64}Cu_{36}$ to $Au_{81}Cu_{19}$ for the Au-Cu IL and of $Au_{83}Cu_{13}Ru_4/Au_{75}Cu_{22}Ru_3$ to $Au_{87}Cu_{11}Ru_2$ for the Au-Cu-Ru IL were measured. For the Au-Cu-Ru IL, the compositions above represent the NPs with the highest and lowest amount of Ru which were detected. Ru appears again to be present only at the surfaces of the NPs obtained from Au-Cu-Ru IL, whereas for the Au-Cu IL binary Au-Cu NPs were identified.

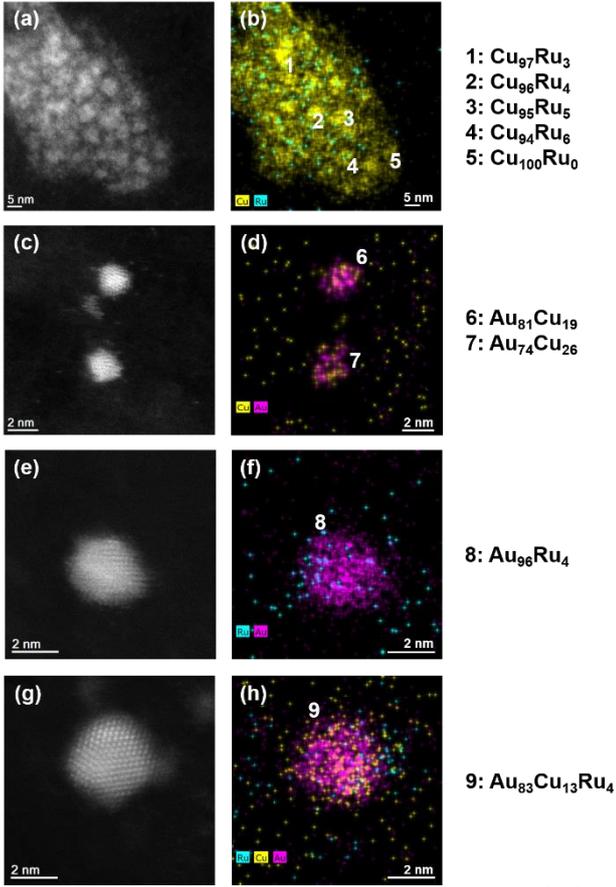

**Figure 4:** HAADF STEM images of co-sputtered NPs are shown in (a) for Cu-Ru, in (c) for Au-Cu, in (e) for Au-Ru and in (f) for Au-Cu-Ru. The elemental distributions of the NPs obtained from the STEM-EDS quantification are depicted in (b) for Cu-Ru, in (d) for Au-Cu, in (f) for Au-Ru and in (h) for Au-Cu-Ru.

### 3.3 Thermodynamic stability from DFT calculations

The structural stability of Au-Cu-Ru and its binary subsystems in different bulk crystal structures was determined by means of the heat of formation $\Delta H$ of a binary phase AB

$$\Delta H = \frac{E_{AB} - N_A E_A - N_B E_B}{N_{AB}} \qquad \text{Equation 1}$$

with respect to the bulk ground states of element $A$ and $B$ using the corresponding total energies $E$ and numbers of atoms $N$. The resulting values of $\Delta H$ of ordered and disordered phases are compiled in Figure 5 for Au-Cu, Au-Ru and Cu-Ru. Negative values of $\Delta H$ indicate alloys that are stable with respect to the decomposition to the elements. In the DFT results, the miscibility of the Au-Cu system manifests in negative values of $\Delta H$ indicating stable $L1_2$-Au$_3$Cu, $L1_0$-AuCu and $L1_2$-AuCu$_3$ phases, in agreement with the experimental phase diagram. The Au-Ru and Cu-Ru systems show positive $\Delta H$ values across the whole composition range confirming the known immiscibility of these systems. The lowest energy structures in these systems, $L1_0$, $L1_2$ and $D0_{22}$, are used to setup the Au-Ru and Cu-Ru NPs (see experimental methods). The results for binary solid solutions are obtained as average of DFT calculations of five SQS structures.

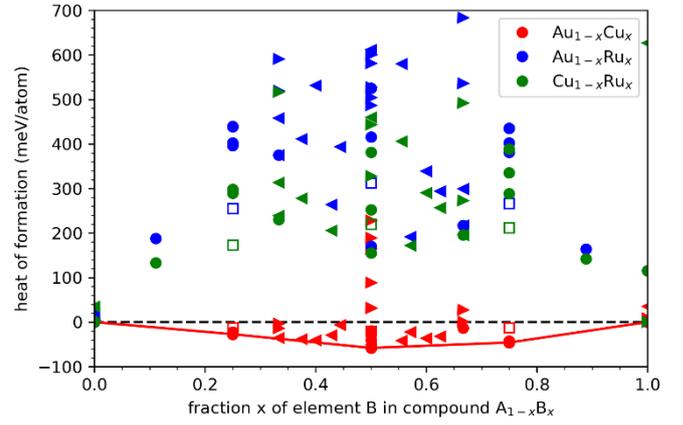

**Figure 5:** Structural stability of binary bulk systems Au-Cu (red), Au-Ru (blue) and Cu-Ru (green) obtained by DFT calculations with respect to the elemental ground states fcc-Au, fcc-Cu and hcp-Ru. The fcc-based, bcc-based and hcp-based ordered structures are indicated as circles, left and right triangles, respectively. Open square symbols represent the average of 5 fcc-based SQS structures with disordered atom arrangement.

For Au-Cu, the considered disordered structures are less preferable than the ordered structures. For Au-Ru and Cu-Ru, the disordered structures are partially energetically more preferable than the ordered counterparts but the positive values of $\Delta H$ rule them thermodynamically unstable, too. Additional calculations based on the thermodynamically stable Au-Cu phases were performed with ternary compositions of 2:1:1 for $L1_0$, 24:7:1 and 23:8:1 for $L1_2$ as well as 6:1:1 for SQS structures. These calculations indicate a small solubility of Ru in Au-Cu but no stable ternary phase with significant Ru content.

The DFT calculations of the NPs use simulation cells of 16 to 140 atoms with Au, Cu, Ru as well as Au-Cu, Au-Ru, and Cu-Ru compositions. The chemical species in the binary NPs are arranged in $L1_0$-AB, $L1_2$-A$_3$B, $L1_2$-AB$_3$, $D0_{22}$-A$_3$B and $D0_{22}$-AB$_3$ orderings that are identified as the energetically most favorable ordered arrangements in the binary systems (Figure 5). Our procedure of constructing the NP by spherical cutouts, motivated by the experimental findings in Figure 3 and 4, leads to very similar initial shapes of the NP of different size (see Figure 6 (a)). For such an isomorphic scaling, the formation energy $\Delta E_{NP}$ of the NPs with respect to the bulk ground-state $E_{bulk}$ (Figure 5) of the same chemical composition can be approximated by a volumetric, a surface and an edge contribution:

$$\Delta E_{NP} = E_{NP} - N_{NP} E_{bulk} = \varepsilon_{vol} V_{NP} + \varepsilon_{surf} V_{NP}^{2/3} + \varepsilon_{edge} V_{NP}^{1/3} \qquad \text{Equation 2}$$

where $E_{NP}$ is the total energy of a NP with $N_{NP}$ atoms and volume $V_{NP} = N_{NP} V_{atom}$ based on the atomic volume $V_{atom}$. Expressing this relation in terms of atoms $N_{NP}$ in the NP,

$$\Delta E_{NP}/N_{NP} = \epsilon_{vol} + \epsilon_{surf} N_{NP}^{-1/3} + \epsilon_{edge} N_{NP}^{-2/3} \qquad \text{Equation 3}$$

gives a simple scaling relation for parameterizing the DFT results of the NP as a function of NP size. The validity of this relation is weakened if the atomic relaxation of the NPs in the DFT calculations distorts the underlying assumption of isomorphic scaling and is hence an indicator of structural changes of the NPs. The fits of the DFT data to the scaling law (Equation 3) presented in the following use the DFT data for the NPs plus additional points for very large $N_{NP}$ where $\Delta E_{NP}$ can be assumed to tend to zero. Figure 6 shows the DFT results and the fits for the unary NPs (in (b)) and for the binary $L1_0$-AB NPs (in (c)). The unary clusters follow very closely the scaling law which validates the applicability of the approach. The DFT results for $L1_0$-AuCu closely follow an isomorphic scaling while $L1_0$-AuRu and $L1_0$-CuRu show sizeable scatter. In the DFT calculations for $L1_2$ and $D0_{22}$ ordering (not shown), the match between DFT results and isomorphic scaling is similar with very good agreement for Au-Cu and sizeable scatter for Au-Ru and Cu-Ru.

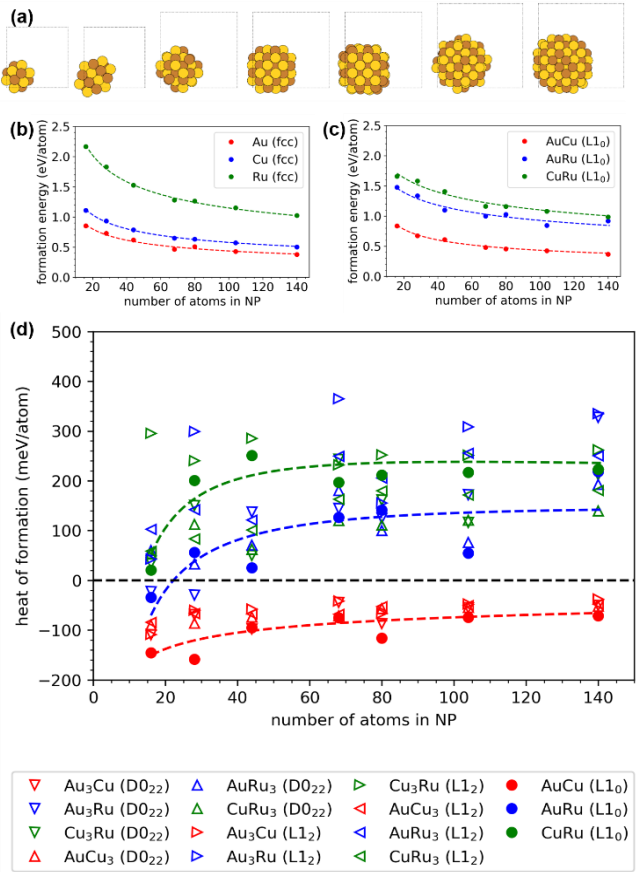

**Figure 6:** DFT calculations were performed for NPs comprising 16 to 140 atoms with different orderings of Au/Cu/Ru. The structures illustrated in (a) are $L1_0$-ordered, relaxed Au-Cu NPs with simulation cells indicated as grey frames. The variations of the formation energy with the number of atoms follow a scaling law with volume, surface and edge contributions (according to Equation 3), as depicted in (b) for unary fcc and in (c) for binary $L1_0$-ordered NPs. The relative formation energies of the different binary NPs with respect to elemental NPs (points: individual DFT calculations, lines: scaling law) in (d) show that only NPs of the miscible Au-Cu system are thermodynamically stable while NPs of the immiscible Au-Ru and Cu-Ru systems are not stable for NPs with more than few 10 atoms.

The relative stability of the binary NPs, i.e. the heat of formation with respect to the unary NPs, is determined in analogy to the bulk structures by Equation 1. Figure 6 (d) shows the size-dependent relative stability of the Au-Cu, Au-Ru and Cu-Ru NPs with $L1_0$-AB, $L1_2$-$A_3B$, $L1_2$-$AB_3$, $D0_{22}$-$A_3B$ and $D0_{22}$-$AB_3$ ordering. The Au-Cu NPs are thermodynamically stable for all considered NP sizes and all Au:Cu ratios with a preference for $L1_0$-AB that is also most stable in the Au-Cu bulk phases. Au-Ru and Cu-Ru NPs are thermodynamically unstable expect for the smallest considered Au-Ru NPs. The formation energy of Ru-containing NPs increases with increasing NP size, as expected from the unstable bulk phases observed in Figure 5. A simple linear interpolation of the formation energy between stable Au-Cu NPs and unstable Au-Ru/Cu-Ru NPs indicates a small solubility of Ru in Au-Cu NPs but no ternary NPs with sizeable Ru content. This is in agreement with the experimentally observed Ru-poor compositions and shows that the bulk miscibility of Au-Cu and the bulk immiscibility of Au-Ru and Cu-Ru is also governing the thermodynamic stability of the NPs. A more detailed understanding of small Ru contents in Au-Cu NPs requires further atomistic simulations in future work.

## 4 Discussion

The co-sputtered thin films served as control samples to verify the capability of solid solution formation of (im-)miscible systems by non-equilibrium sputtering onto solid substrates. Comparing the composition and structure of those thin films with co-sputtered NPs/ILs could answer whether sputtering onto ILs corresponds also to non-equilibrium processes feasible of alloying immiscible elements in solid solution NPs.

### 4.1 Analysis of the co-deposited thin films

The DFT calculations confirm the immiscibility of Ru with Au and Cu as expected by the Hume-Rothery rules due to the different crystal lattices (Ru: hcp vs. Au/Cu: fcc). The known Au-Cu bulk phases ($Au_3Cu$, AuCu and $AuCu_3$) form the convex hull with negative values of the heat of formation (see Figure 5). The Au-Ru and Cu-Ru systems, in contrast, show no stable intermetallic phase across the range of chemical composition. The solubility of Ru in Au-Cu is very low and therefore Au-Cu-Ru bulk alloys with sizeable Ru content are not expected to be thermodynamically stable.

However, the EDS results in Table 3 and the XRD patterns in Figure 1 contradict these general principles. The thin films show a nearly equiatomic composition and the XRD patterns confirm the existence of solid solutions in all thin films. For Au-Cu, this is not surprising due to the complete solid solubility of Au and Cu and since AuCu is an intermetallic phase with space group *P*4/*mmm* and *Imma*.[69] The compositional deviation of 3 at.% between the co-sputtered Au-Cu thin film ($Au_{47}Cu_{53}$) and AuCu results in the slight shift of the XRD peaks from the literature values according to Vegard's law and thus indicates that the co-sputtered thin film is a solid solution.

For Cu-Ru and Au-Ru thin films, the observed fcc structure is surprising, since the formation of the Cu-Ru and Au-Ru solid solutions showing also fcc structure contradicts the DFT results. However, similar observations were made for equiatomic CuRu[66] and AuRu NPs[68] as well as for $AuRu_3$[44] NPs, showing also fcc structure. The linear change of the lattice constants with the NP compositions according to Vegard's law and the homogeneous distribution of the elements in the binary NPs according to STEM-EDS maps indicate the existence of a solid solution for Cu-Ru and Au-Ru. Since also pure Ru NPs with fcc structure were synthesized,[70] there is no doubt that in the thin films solid solutions with fcc structure of the bulk immiscible elements Au-Ru and Cu-Ru exist. For the Au-Cu-Ru thin film, the match between the XRD peaks and the AuCu reference data (fcc structure) also supports the assumption that Ru is incorporated into the fcc solid solution. This confirms the ability of incorporating bulk-immiscible elements into thin film solid solutions by sputtering on solid substrates without heating, which is a non-equilibrium process. The sputtered atoms arrive at the substrate which is close to room temperature as it is not actively heated. Thus, the condensed atoms do not have sufficient energy to diffuse over the surface and are buried below the further arriving sputtered atoms. Since the deposition rates were chosen to obtain an equiatomic composition, a homogeneous binary (and ternary) atom mixture on the surface is achieved. The sputtered atoms are quenched into a forced solid solution and no decomposition into the respective unary phases occurs since no energy is provided for developing an energetic most stable, demixed ordering, typical for a non-equilibrium process.

### 4.1 Analysis of the co-sputtered nanoparticles

The results of the co-sputtered thin films suggest that immiscible elements could be combined into solid-solution NPs by co-sputtering on ILs. According to the ICP-MS results (Table 4), the composition of the ILs in terms of the co-sputtered elements is in good agreement with the thin films (see Table 3), so that the conditions for solid solution formation should be given according to the compositions of the thin films showing such solid solutions.

However, the STEM-EDS investigations of the co-sputtered NPs in Figure 4 reveal that no incorporation of Ru in the NPs obtained from the Cu-Ru, the Au-Ru and the Au-Cu-Ru deposition occurred. On the other hand, binary Au-Cu NPs were identified in the Au-Cu and the Au-Cu-Ru NP/IL-suspensions. Ru atoms in the vicinity of the NPs appear to sit only at the surface of the individual NPs without being incorporated into the crystal structure or building a closed shell. These assumptions are underlined by the composition of individual NPs, revealing mainly compositions of $Cu_{94}Ru_6$ to $Cu_{100}Ru_0$ for the Cu-Ru IL, of $Au_{96}Ru_4$ to $Au_{99}Ru_1$ for the Au-Ru IL, of $Au_{64}Cu_{36}$ to $Au_{81}Cu_{19}$ for the Au-Cu IL and of $Au_{83}Cu_{13}Ru_4/Au_{75}Cu_{22}Ru_3$ to $Au_{87}Cu_{11}Ru_2$ for the Au-Cu-Ru IL, as determined by the quantification of the STEM-EDS measurements. For the Au-Cu-Ru IL, the NP composition represents the highest and the lowest detected amount of Ru for individual NPs.

The mean NP diameters and the general size distributions of the co-sputtered NPs also indicate a difference between the ILs containing only bulk-immiscible elements and the ILs containing bulk-miscible elements (see Figure 2). The comparison of the results shows that the NPs in the Au-Cu and the Au-Cu-Ru IL (the ILs containing bulk-miscible elements) have nearly identical mean diameters (Au-Cu: $(1.8 \pm 0.6)$ nm, Au-Cu-Ru: $(1.7 \pm 0.7)$ nm). Additionally, the size distributions for the Au-Cu and the Au-Cu-Ru IL show a nearly identical width and only a slight difference in the maximum NP diameter (3 nm for Au-Cu IL and 4 nm for Au-Cu-Ru IL). For the Cu-Ru and Au-Ru ILs, containing only bulk-immiscible elements, the NP mean diameters of $(5.0 \pm 1.7)$ nm for Cu-Ru IL and $(3.3 \pm 1.4)$ nm for Au-Ru IL are much bigger than the mean diameters for the ILs containing miscible elements. This goes along with much broader size distributions and higher maximum NP diameters (9 nm for Cu-Ru IL, 6 nm for Au-Ru IL) for the ILs with only bulk-immiscible elements with respect to the ILs containing bulk-miscible elements. This could be explained by the following assumption: The Cu-Ru and the Au-Ru IL contain only pure Cu and Au NPs respectively, decorated with Ru atoms at their surface but without a complete closed Ru outer shell. The Au-Cu IL and the Au-Cu-Ru IL contain Au-Cu alloy NPs with also a Ru decoration on the outer shell in the case of the Au-Cu-Ru IL.

A further comparison of those results concerning mean diameters and size distributions with the same NP characteristics of pure Au, Cu and binary Au-Cu NPs also obtained from (co-)sputtering for a former publication[65] supports this assumption. This publication also addresses the size differences between unary and binary NP/IL-systems. The size distributions of the Au-Cu and the Au-Cu-Ru IL show a comparable narrowness with respect to the size distribution of the Au-Cu IL obtained previously.[65] The size distribution of the Cu-Ru and the Au-Ru IL are also very comparable to the size distribution of the NPs in pure Cu and Au IL from the previous publication[65] in terms of their width. However, the NP mean diameters of the Au-Ru and Cu-Ru IL are much bigger than the NP mean diameters of pure Au and Cu IL and the size distributions of the binary ILs are shifted more towards bigger NP diameters. This may be attributed to a higher temperature of the substrate plate holding the Cu-Ru IL and the Au-Ru IL during the co-deposition since two sputter fluxes arrive at the plate. In the previous publication, annealing of the sputtered ILs also resulted in increasing NP diameters.[65]

Thus, the results of the STEM-EDS measurements and the comparison of the size distributions emphasize the formation of pure Au NPs in the Au-Ru IL, of pure Cu NPs in the Cu-Ru IL and of binary Au-Cu NPs in the Au-Cu IL and the Au-Cu-Ru IL, all of them decorated with Ru at the NP surfaces (with Ru present in the IL). The arising question is, why the bulk-immiscible elements could not be combined into alloy NPs by sputtering onto ILs while they could be transferred into solid solutions with good crystallinity when sputtered onto solid substrates. A key for answering this question is the formation process of the NPs in ILs. It is still discussed where NP nucleation occurs, either at the IL surface[71,72] or in the IL volume[73,74]. However, our recent investigation comprising reactive oxygen deposition of Cu onto the surface of two different ILs with different surface tensions and viscosities shows strong indications that the NP formation occurs within the IL volume after the sputtered atoms penetrated into the IL without spending time at the surface.[51]

A schematic in Figure 7 shows the assumed NP formation process when sputtering onto ILs for the co-sputtered systems Cu-Ru in (a), Au-Cu in (b), Au-Ru in (c), Au-Cu-Ru in (d) and for the hypothesis of the NP formation occurring at the IL surface in (e). The sputtered atoms penetrate directly into the IL because of their high kinetic energy. Due to their small size, individual sputtered atoms are too small to be influenced or stabilized by the IL ions.[51] Atom enrichment in the near surface region (i) of the IL results in collisions of the sputtered atoms and the formation of first atom clusters. Collisions of atoms with the primary atom clusters (ii) result in growth of theses clusters, which grow into NPs. The NPs disperse further into the IL volume (iii) when they become bigger and thus heavier and are finally stabilized by the IL ions when they descent out of the region with increased sputtered atom concentration. This formation and growth process was deduced from our results of reactive oxygen sputtering of Cu onto the surfaces of the ILs [Bmim][(Tf)$_2$N] (used also here) and 1-etyl-3-methylimidazolium bis-(trifluoromethylsulfonyl) imide [Emim][(Tf)$_2$N].[51] It corresponds to an equilibrium process since the sputtered atoms can move freely in the IL volume and do not experience any external driving force which could quench them into alloy NPs during NP formation.

If the NPs would form at the IL surface, the sputtered atoms would stay a certain time at the surface before immersing into the IL volume. During this time, the same processes like on a solid substrate would arise: alloying of bulk-immiscible elements into solid solution NPs would occur due to the sputtered atoms being buried by the further incoming atoms of all sputtered elements before the NPs immerse into the IL volume and become stabilized by the IL. Since the composition of the sputter flux on the IL surface is identical to the flux arriving at the added Si/SiO$_2$ solid substrate, the composition of the NPs should be comparable to the thin film composition with Ru being present and uniformly distributed in all NPs. Since the measurement data contradict this scenario, the assumption of the NP formation occurring at the IL surface under non-equilibrium conditions is again disproved. In the case of the NP formation in ILs being an equilibrium process, the statements of the Hume-Rothery rules for estimating the probability of alloy formation are important,[75] with the modification that the part concerning a similar electronegativity should be replaced by a rule based on the molar heat of vaporization related to the cohesive energy of materials.[65,75–77] This modification results in a better applicability of the Hume-Rothery rules for the alloy formation in the nanoscale.[75] For the Cu-Ru, Au-Cu, Au-Ru and Au-Cu-Ru systems, the relevant elemental parameters are listed in Table 1. For all three elements, the difference in the atomic radius is < 15% and all elements show a common valence (Au and Cu both 1+, Au and Ru both 3+ and Cu and Ru both 2+), so that two Hume-Rothery rules are fulfilled. However, Ru crystallizes in the hcp structure whereas the fcc structure occurs for Au and Cu. Furthermore, the molar heat of vaporization of Ru is nearly twice the value for Au and Cu, conflicting with the demand for a good accordance of this parameter for both alloyed elements. This explains that only Au and Cu can form an alloy according to the Hume-Rothery rules and that Ru cannot be alloyed with Au or Cu in an equilibrium process. These expectations from the Hume-Rothery rules are confirmed in the presented DFT simulations of the thermodynamic stability of different NPs. The bulk-miscible combination Au-Cu shows isomorphic scaling of the relative formation energy across the investigated range of NP sizes and thermodynamically stable NPs for the considered compositions $Au_3Cu$, $AuCu$ and $AuCu_3$. The bulk-immiscible combinations Au-Ru and Cu-Ru, however, do not form thermodynamically stable NPs for sizes that are comparable to our samples (at least 100 atoms). For these two systems, the formation of pure Au, pure Cu and pure Ru NPs is thermodynamically preferable. For all considered Au-Ru and Cu-Ru compositions, the isomorphic scaling is less pronounced for larger NPs, which may indicate considerable structural deformations. The theoretical solubility of Ru in Au-Cu structures does not indicate that Au-Cu-Ru alloys can be synthesized. The amount of solvable Ru in Au-Cu is so low that this system cannot be considered a ternary alloy. For all systems in all compositions, it is shown that the (im-)miscibility of the bulk systems also holds for the NPs.

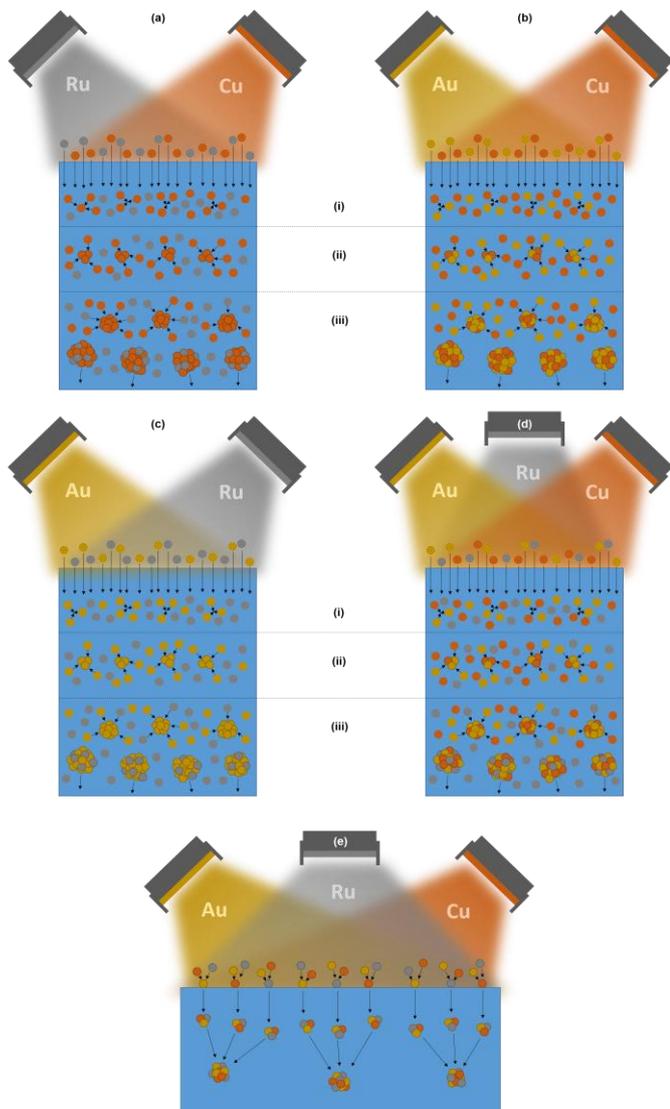

**Figure 7:** Schematics of NP formation mechanisms in ILs obtained from co-sputter deposition processes of elements with different (im-)miscibility. The rotation of the substrate results in a homogeneous distribution of the elements over the whole IL surface. The formation of alloy NPs depends on the co-sputtered elements: For Cu-Ru in (a) and Au-Ru in (c), the different characteristics of the bulk-immiscible elements result in pure Cu (a) and Au (c) NPs with Ru atoms present in the IL and as impurities at the NP surfaces. For Au-Cu in (b) an alloy can form, resulting in Au-Cu NPs. For the co-deposition of Au-Cu-Ru in (d), the bulk-miscible Au and Cu atoms form Au-Cu NPs with the Ru remaining in the IL and being present also at the NP surfaces. Image (e) shows the NP formation with the assumption that it occurs at the IL surface. The sputtered atoms grow at the IL surface into primary atom clusters with homogeneous composition and grow further due to the sputter flux before they immerse into the IL volume and become stabilized by the IL ions. In the IL volume, further NP growth due to collisions and coalescence of the primary atom clusters occurs. This formation process is **not** the process we assume to occur on the basis of the obtained experimental results.

This means that for the bulk-immiscible elements Cu-Ru and Au-Ru only pure Cu NPs and Au NPs form, as depicted in Figure 7 (a) and (c). Ru atoms are attached to a small degree on the NP surfaces but not in the NP volume and thus mainly stay in the IL, which means they are removed with the IL during TEM grid preparation. For the bulk-miscible elements Au and Cu, Au-Cu alloy NPs can form, as depicted in Figure 7 (b). Co-sputtering of three elements results in the formation of Au-Cu alloy NPs decorated with Ru atoms at the NP surfaces, illustrated in Figure 7 (d). The excess of sputtered Ru atoms stays in the IL and is removed during TEM grid preparation.

The specific NP compositions for the bulk-miscible and bulk-immiscible element pairs show an interesting coincidence. Besides the missing incorporation of Ru in the NPs obtained from Cu-Ru and Au-Ru co-sputtering, the Au-Cu NPs in the Au-Cu IL and the Au-Cu-Ru IL show compositions from $Au_{64}Cu_{36}$ to $Au_{89}Cu_{11}$, which is in good agreement with the most stable composition of $Au_{86}Cu_{14}$ and $Au_{87}Cu_{13}$ as determined by genetic algorithm calculations for Au-Cu bimetallic clusters.[65,78,79] Since the most stable NP composition is formed by Au-Cu NPs obtained from co-sputtering, the assumption of a usual equilibrium process being responsible for the NP formation when co-sputtering on ILs is further supported.

Au-Ru[44,68] and Cu-Ru[42,66,80,81] NPs with a uniform distribution of the elements in the NPs (no core-shell) reported in literature have been synthesized with a reduction technique using reaction conditions differing from ambient conditions or precursors with suitable structural accordance and modifications and similar reaction times to achieve a better miscibility of both elements. Those techniques are closer to a non-equilibrium process due to the non-ambient reaction conditions and the selected precursors than with an equilibrium process.

Thus, a non-equilibrium process is necessary to alloy the bulk-immiscible elements in NPs stabilized in ILs, as demonstrated by our measurements, the DFT results and the literature and as supported by the above-presented model for the NP formation when sputtering onto ILs. In contrast to that, sputtering on ILs for the NP synthesis has characteristics of an equilibrium process due to the NP formation process in ILs.

# 5 Conclusion

Our experiments confirm that sputtering can quench bulk-immiscible elements into solid solutions on an unheated solid substrate and thus corresponds to a non-equilibrium process. The bulk-immiscible element combinations Cu-Ru, Au-Ru and Au-Cu-Ru were synthesized as solid solutions on a solid substrate, but failed in forming solid solution NPs when sputtered on ILs. DFT calculations of the thermodynamic stability for the binary and ternary bulk systems and nanoparticles confirm those results which can be explained by the unique NP formation process in ILs. The sputtered atoms penetrate directly into the IL volume and move there freely in the IL matrix. The NP formation is induced by coalescence of the free "swimming" atoms according to a usual equilibrium process without external driving force quenching the bulk-immiscible elements into NPs. Thus, the Hume-Rothery rules are valid with the modification of considering the molar heat of vaporization as replacement for electronegativity. These modified rules show a better applicability for processes in the nanoscale and explain also our findings of only bulk-miscible binary or unary Au and Cu NPs in the ILs when sputtering the Au-Cu-Ru system combinations onto ILs.

# Author Contributions

The manuscript was written by M. Meischein except for the parts concerning the DFT calculations in the Experimental data, Discussion and Experimental methods sections, written by T. Hammerschmidt. Microscopy data were provided by M. Meischein (conventional TEM) and A. Garzón-Manjón, S. Zhang and L. Abdellaoui (HAADF-STEM and STEM-EDS). XRD data were provided by B. Xiao. DFT simulations were conducted by T. Hammerschmidt. C. Scheu and A. Ludwig provided funding as well as scientific guidance for the project and co-wrote the manuscript.

# Conflicts of interest

There are no conflicts to declare.

# Acknowledgements

This work was funded by the German Science Foundation (DFG) via the project SCHE 634/21-1 and LU 1175/23-1. We want to thank Mr. Quan Zhang from the group of Hiroshi Kitagawa at the Kyoto University for providing XRD data of the Au-Ru NPs, which were measured with a conventional XRD system with Cu $K_\alpha$ radiation, for comparison with our Au-Ru thin films. Moreover, we want to thank the Center of Electrochemical Sciences (CES) at the Chair for Analytical Chemistry of the Faculty of Chemistry and Biochemistry at Ruhr University Bochum, for conducting the ICP-MS measurements operated by Martin Trautmann.

# Notes and references

1   R. Weissleder, G. Elizondo, J. Wittenberg, C. A. Rabito, H. H. Bengele and L. Josephson, Ultrasmall superparamagnetic iron oxide: characterization of a new class of contrast agents for MR imaging, *Radiology*, 1990, **175**, 489–493.

2   M. Breisch, V. Grasmik, K. Loza, K. Pappert, A. Rostek, N. Ziegler, A. Ludwig, M. Heggen, M. Epple, J. C. Tiller, T. A. Schildhauer, M. Köller and C. Sengstock, Bimetallic silver-platinum nanoparticles with combined osteo-promotive and antimicrobial activity, *Nanotechnology*, 2019, **30**, 305101.

3   Q. A. Pankhurst, J. Connolly, S. K. Jones and J. Dobson, Applications of magnetic nanoparticles in biomedicine, *J. Phys. D: Appl. Phys.*, 2003, **36**, R167-R181.

4   M. Shinkai, M. Yanase, M. Suzuki, H. Honda, T. Wakabayashi, J. Yoshida and T. Kobayashi, Intracellular hyperthermia for cancer using magnetite cationic liposomes, *Journal of Magnetism and Magnetic Materials*, 1999, **194**, 176–184.

5   J. Li and J. Z. Zhang, Optical properties and applications of hybrid semiconductor nanomaterials, *Coordination Chemistry Reviews*, 2009, **253**, 3015–3041.

6   S. Karmakar, S. Kumar, R. Rinaldi and G. Maruccio, Nano-electronics and spintronics with nanoparticles, *J. Phys.: Conf. Ser.*, 2011, **292**, 12002.

7   D. Astruc, *Nanoparticles and catalysis*, Wiley-VCH, Weinheim, 2008.

8   P. Migowski and J. Dupont, Catalytic applications of metal nanoparticles in imidazolium ionic liquids, *Chemistry (Weinheim an der Bergstrasse, Germany)*, 2007, **13**, 32–39.

9   T. Löffler, A. Savan, A. Garzón-Manjón, M. Meischein, C. Scheu, A. Ludwig and W. Schuhmann, Toward a Paradigm Shift in Electrocatalysis Using Complex Solid Solution Nanoparticles, *ACS Energy Lett.*, 2019, **4**, 1206–1214.

10  T. Löffler, H. Meyer, A. Savan, P. Wilde, A. Garzón Manjón, Y.-T. Chen, E. Ventosa, C. Scheu, A. Ludwig and W.

Schuhmann, Discovery of a Multinary Noble Metal-Free Oxygen Reduction Catalyst, *Adv. Energy Mater.*, 2018, **8**, 1802269.

11   Y. Yao, Z. Huang, P. Xie, S. D. Lacey, R. J. Jacob, H. Xie, F. Chen, A. Nie, T. Pu, M. Rehwoldt, D. Yu, M. R. Zachariah, C. Wang, R. Shahbazian-Yassar, J. Li and L. Hu, Carbothermal shock synthesis of high-entropy-alloy nanoparticles, *Science (New York, N.Y.)*, 2018, **359**, 1489–1494.

12   C. Yang, B. H. Ko, S. Hwang, Z. Liu, Y. Yao, W. Luc, M. Cui, A. S. Malkani, T. Li, X. Wang, J. Dai, B. Xu, G. Wang, D. Su, F. Jiao and L. Hu, Overcoming immiscibility toward bimetallic catalyst library, *Science advances*, 2020, **6**, eaaz6844.

13   K. Kusada, D. Wu and H. Kitagawa, New Aspects of Platinum Group Metal-Based Solid-Solution Alloy Nanoparticles: Binary to High-Entropy Alloys, *Chemistry (Weinheim an der Bergstrasse, Germany)*, 2020, **26**, 5105–5130.

14   D. König, K. Richter, A. Siegel, A.-V. Mudring and A. Ludwig, High-Throughput Fabrication of Au-Cu Nanoparticle Libraries by Combinatorial Sputtering in Ionic Liquids, *Adv. Funct. Mater.*, 2014, **24**, 2049–2056.

15   M. Hirano, K. Enokida, K.-i. Okazaki, S. Kuwabata, H. Yoshida and T. Torimoto, Composition-dependent electrocatalytic activity of AuPd alloy nanoparticles prepared via simultaneous sputter deposition into an ionic liquid, *Physical chemistry chemical physics: PCCP*, 2013, **15**, 7286–7294.

16   E. Redel, R. Thomann and C. Janiak, First correlation of nanoparticle size-dependent formation with the ionic liquid anion molecular volume, *Inorganic chemistry*, 2008, **47**, 14–16.

17   E. Redel, R. Thomann and C. Janiak, Use of ionic liquids (ILs) for the IL-anion size-dependent formation of Cr, Mo and W nanoparticles from metal carbonyl M(CO)6 precursors, *Chemical communications (Cambridge, England)*, 2008, 1789–1791.

18   P. Migowski, G. Machado, S. R. Texeira, M. C. M. Alves, J. Morais, A. Traverse and J. Dupont, Synthesis and characterization of nickel nanoparticles dispersed in imidazolium ionic liquids, *Physical chemistry chemical physics: PCCP*, 2007, **9**, 4814–4821.

19   Z. He and P. Alexandridis, Nanoparticles in ionic liquids: interactions and organization, *Physical chemistry chemical physics: PCCP*, 2015, **17**, 18238–18261.

20   M. Meischein, M. Fork and A. Ludwig, On the Effects of Diluted and Mixed Ionic Liquids as Liquid Substrates for the Sputter Synthesis of Nanoparticles, *Nanomaterials (Basel, Switzerland)*, 2020, **10**. DOI: 10.3390/nano10030525.

21   J. D. Holbrey and K. R. Seddon, Ionic Liquids, *Clean Technologies and Environmental Policy*, 1999, **1**, 223–236.

22   P. Wasserscheid and W. Keim, Ionic Liquids—New "Solutions" for Transition Metal Catalysis, *Angew. Chem. Int. Ed.*, 2000, **39**, 3772–3789.

23   S. Kuwabata, T. Torimoto, A. Imanishi and T. Tsu, in *Ionic Liquids - New Aspects for the Future*, ed. J.-i. Kadokawa, InTech, 2013.

24   R. L. Vekariya, A review of ionic liquids: Applications towards catalytic organic transformations, *Journal of Molecular Liquids*, 2017, **227**, 44–60.

25   N. V. Plechkova and K. R. Seddon, Applications of ionic liquids in the chemical industry, *Chemical Society reviews*, 2008, **37**, 123–150.

26   P. Wasserscheid and T. Welton, *Ionic Liquids in Synthesis*, Wiley-VCH, Hoboken, 2nd edn., 2008.

27   K. L. Luska, P. Migowski and W. Leitner, Ionic liquid-stabilized nanoparticles as catalysts for the conversion of biomass, *Green Chem.*, 2015, **17**, 3195–3206.

28   F. M. Kerton and R. Marriott, *Alternative solvents for green chemistry*, RSC Publ, Cambridge, 2nd edn., 2013.

29  B. Hashemi, P. Zohrabi and S. Dehdashtian, Application of green solvents as sorbent modifiers in sorptive-based extraction techniques for extraction of environmental pollutants, *TrAC Trends in Analytical Chemistry*, 2018, **109**, 50–61.

30  H. Okamoto, D. J. Chakrabarti, D. E. Laughlin and T. B. Massalski, The Au−Cu (Gold-Copper) system, *JPE*, 1987, **8**. DOI: 10.1007/BF02893155.

31  H. Okamoto and T. B. Massalski, The Au-Ru (Gold-Ruthenium) system, *Bulletin of Alloy Phase Diagrams*, 1984, **5**, 388–390.

32  U. Mizutani, Hume-Rothery rules for structurally complex alloy phases, *MRS Bull.*, 2012, **37**, 169.

33  W. Hume-Rothery, Mabbott, Gilbert, W. and K. M. Channel Evans, The freezing points, melting points, and solid solubility limits of the alloys of sliver and copper with the elements of the b sub-groups, *Phil. Trans. R. Soc. Lond. A*, 1934, **233**, 1–97.

34  W. Martienssen and H. Warlimont, *Springer Handbook of Condensed Matter and Materials Data*, Springer Berlin Heidelberg, 2005.

35  M. Winter, WebElements: THE periodic table on the WWW, https://www.webelements.com/, (accessed 7th March 2022).

36  A. Jain, S. P. Ong, G. Hautier, W. Chen, W. D. Richards, S. Dacek, S. Cholia, D. Gunter, D. Skinner, G. Ceder and K. A. Persson, Commentary: The Materials Project: A materials genome approach to accelerating materials innovation, *APL Materials*, 2013, **1**, 11002.

37  M. de Jong, W. Chen, T. Angsten, A. Jain, R. Notestine, A. Gamst, M. Sluiter, C. Krishna Ande, S. van der Zwaag, J. J. Plata, C. Toher, S. Curtarolo, G. Ceder, K. A. Persson and M. Asta, Charting the complete elastic properties of inorganic crystalline compounds, *Scientific data*, 2015, **2**, 150009.

38  X. Liu, A. Wang, X. Wang, C.-Y. Mou and T. Zhang, Au-Cu Alloy nanoparticles confined in SBA-15 as a highly efficient catalyst for CO oxidation, *Chemical communications (Cambridge, England)*, 2008, 3187–3189.

39  J. Llorca, M. Dominguez, C. Ledesma, R. Chimentao, F. Medina, J. Sueiras, I. Angurell, M. Seco and O. Rossell, Propene epoxidation over TiO2-supported Au–Cu alloy catalysts prepared from thiol-capped nanoparticles, *Journal of Catalysis*, 2008, **258**, 187–198.

40  R. J. Chimentão, F. Medina, J.L.G. Fierro, J. Llorca, J. E. Sueiras, Y. Cesteros and P. Salagre, Propene epoxidation by nitrous oxide over Au–Cu/TiO2 alloy catalysts, *Journal of Molecular Catalysis A: Chemical*, 2007, **274**, 159–168.

41  J. Liu, L. L. Zhang, J. Zhang, T. Liu and X. S. Zhao, Bimetallic ruthenium-copper nanoparticles embedded in mesoporous carbon as an effective hydrogenation catalyst, *Nanoscale*, 2013, **5**, 11044–11050.

42  B. Huang, H. Kobayashi, T. Yamamoto, S. Matsumura, Y. Nishida, K. Sato, K. Nagaoka, S. Kawaguchi, Y. Kubota and H. Kitagawa, Solid-Solution Alloying of Immiscible Ru and Cu with Enhanced CO Oxidation Activity, *Journal of the American Chemical Society*, 2017, **139**, 4643–4646.

43  P. S. S. Kumar, A. Manivel, S. Anandan, M. Zhou, F. Grieser and M. Ashokkumar, Sonochemical synthesis and characterization of gold–ruthenium bimetallic nanoparticles, *Colloids and Surfaces A: Physicochemical and Engineering Aspects*, 2010, **356**, 140–144.

44  Q. Zhang, K. Kusada, D. Wu, T. Yamamoto, T. Toriyama, S. Matsumura, S. Kawaguchi, Y. Kubota and H. Kitagawa, Selective control of fcc and hcp crystal structures in Au-Ru solid-solution alloy nanoparticles, *Nature communications*, 2018, **9**, 510.

45  W. G. J. Bunk, *Advanced Structural and Functional Materials*, Springer Berlin Heidelberg, Berlin, Heidelberg, 1991.

46  G. Schmid, Clusters and colloids: bridges between molecular and condensed material, *Endeavour*, 1990, **14**, 172–178.

47 K. T. Ramesh, *Nanomaterials*, Springer US, Boston, MA, 2009.

48 M. Zehetbauer and Y. T. Zhu, *Bulk nanostructured materials*, Wiley-VCH, Weinheim, 2009.

49 Y. Yan, J. S. Du, K. D. Gilroy, D. Yang, Y. Xia and H. Zhang, Intermetallic Nanocrystals: Syntheses and Catalytic Applications, *Advanced materials (Deerfield Beach, Fla.)*, 2017, **29**. DOI: 10.1002/adma.201605997.

50 Y. J. Li, A. Savan, A. Kostka, H. S. Stein and A. Ludwig, Accelerated atomic-scale exploration of phase evolution in compositionally complex materials, *Mater. Horiz.*, 2018, **5**, 86–92.

51 M. Meischein, X. Wang and A. Ludwig, Unraveling the Formation Mechanism of Nanoparticles Sputtered in Ionic Liquid, *J. Phys. Chem. C*, 2021, **125**, 24229–24239.

52 H. Meyer, M. Meischein and A. Ludwig, Rapid Assessment of Sputtered Nanoparticle Ionic Liquid Combinations, *ACS combinatorial science*, 2018, **20**, 243–250.

53 K. Momma and F. Izumi, VESTA 3 for three-dimensional visualization of crystal, volumetric and morphology data, *J Appl Crystallogr*, 2011, **44**, 1272–1276.

54 S. Zhang and C. Scheu, Evaluation of EELS spectrum imaging data by spectral components and factors from multivariate analysis, *Microscopy (Oxford, England)*, 2018, **67**, i133-i141.

55 G. Kresse and J. Hafner, Ab initio molecular-dynamics simulation of the liquid-metal–amorphous-semiconductor transition in germanium, *Physical review. B, Condensed matter*, 1994, **49**, 14251–14269.

56 G. Kresse and J. Furthmüller, Efficiency of ab-initio total energy calculations for metals and semiconductors using a plane-wave basis set, *Computational Materials Science*, 1996, **6**, 15–50.

57 G. Kresse and J. Furthmüller, Efficient iterative schemes for ab initio total-energy calculations using a plane-wave basis set, *Physical review. B, Condensed matter*, 1996, **54**, 11169–11186.

58 T. Hammerschmidt, A. F. Bialon, D. G. Pettifor and R. Drautz, Topologically close-packed phases in binary transition-metal compounds: matching high-throughput ab initio calculations to an empirical structure map, *New J. Phys.*, 2013, **15**, 115016.

59 P. E. Blöchl, Projector augmented-wave method, *Physical review. B, Condensed matter*, 1994, **50**, 17953–17979.

60 J. P. Perdew, K. Burke and M. Ernzerhof, Generalized Gradient Approximation Made Simple, *Physical review letters*, 1996, **77**, 3865–3868.

61 H. J. Monkhorst and J. D. Pack, Special points for Brillouin-zone integrations, *Phys. Rev. B*, 1976, **13**, 5188–5192.

62 B. Grabowski, T. Hickel and J. Neugebauer, Ab initio study of the thermodynamic properties of nonmagnetic elementary fcc metals: Exchange-correlation-related error bars and chemical trends, *Phys. Rev. B*, 2007, **76**. DOI: 10.1103/PhysRevB.76.024309.

63 A. Zunger, S.-H. Wei, L. G. Ferreira and J. E. Bernard, Special quasirandom structures, *Physical review letters*, 1990, **65**, 353–356.

64 A. Walle and G. Ceder, Automating first-principles phase diagram calculations, *JPE*, 2002, **23**, 348–359.

65 M. Meischein, A. Garzón-Manjón, T. Frohn, H. Meyer, S. Salomon, C. Scheu and A. Ludwig, Combinatorial Synthesis of Binary Nanoparticles in Ionic Liquids by Cosputtering and Mixing of Elemental Nanoparticles, *ACS combinatorial science*, 2019, **21**, 743–752.

66 B. Huang, H. Kobayashi, T. Yamamoto, T. Toriyama, S. Matsumura, Y. Nishida, K. Sato, K. Nagaoka, M. Haneda, W. Xie, Y. Nanba, M. Koyama, F. Wang, S. Kawaguchi, Y. Kubota and H. Kitagawa, A CO Adsorption Site Change Induced by Copper Substitution in a Ruthenium Catalyst for Enhanced CO Oxidation Activity, *Angew. Chem. Int. Ed.*, 2019, **58**, 2230–2235.

67  I. Uszyński, J. Janczak and R. Kubiak, Thermal expansion of α-AuCu, AuCu(II) and AuCu(I) at low temperatures, *Journal of Alloys and Compounds*, 1994, **206**, 211–213.

68  Q. Zhang, K. Kusada, D. Wu, N. Ogiwara, T. Yamamoto, T. Toriyama, S. Matsumura, S. Kawaguchi, Y. Kubota, T. Honma and H. Kitagawa, Solid-solution alloy nanoparticles of a combination of immiscible Au and Ru with a large gap of reduction potential and their enhanced oxygen evolution reaction performance, *Chemical science*, 2019, **10**, 5133–5137.

69  C. L. Bracey, P. R. Ellis and G. J. Hutchings, Application of copper-gold alloys in catalysis: current status and future perspectives, *Chemical Society reviews*, 2009, **38**, 2231–2243.

70  K. Kusada, H. Kobayashi, T. Yamamoto, S. Matsumura, N. Sumi, K. Sato, K. Nagaoka, Y. Kubota and H. Kitagawa, Discovery of face-centered-cubic ruthenium nanoparticles: facile size-controlled synthesis using the chemical reduction method, *Journal of the American Chemical Society*, 2013, **135**, 5493–5496.

71  Y. Hatakeyama, M. Okamoto, T. Torimoto, S. Kuwabata and K. Nishikawa, Small-Angle X-ray Scattering Study of Au Nanoparticles Dispersed in the Ionic Liquids 1-Alkyl-3-methylimidazolium Tetrafluoroborate, *J. Phys. Chem. C*, 2009, **113**, 3917–3922.

72  H. Wender, L. F. de Oliveira, P. Migowski, A. F. Feil, E. Lissner, M. H. G. Prechtl, S. R. Teixeira and J. Dupont, Ionic Liquid Surface Composition Controls the Size of Gold Nanoparticles Prepared by Sputtering Deposition, *J. Phys. Chem. C*, 2010, **114**, 11764–11768.

73  Y. Hatakeyama, K. Onishi and K. Nishikawa, Effects of sputtering conditions on formation of gold nanoparticles in sputter deposition technique, *RSC Adv.*, 2011, **1**, 1815.

74  E. Vanecht, K. Binnemans, J. W. Seo, L. Stappers and J. Fransaer, Growth of sputter-deposited gold nanoparticles in ionic liquids, *Physical chemistry chemical physics: PCCP*, 2011, **13**, 13565–13571.

75  G. Guisbiers, S. Mejia-Rosales, S. Khanal, F. Ruiz-Zepeda, R. L. Whetten and M. José-Yacaman, Gold-copper nano-alloy, "Tumbaga", in the era of nano: phase diagram and segregation, *Nano letters*, 2014, **14**, 6718–6726.

76  K. Miyajima, N. Fukushima, H. Himeno, A. Yamada and F. Mafuné, Breakdown of the Hume-Rothery rules in sub-nanometer-sized Ta-containing bimetallic small clusters, *The journal of physical chemistry. A*, 2009, **113**, 13448–13450.

77  K. Miyajima, H. Himeno, A. Yamada, H. Yamamoto and F. Mafuné, Nanoalloy formation of Ta-containing trimetallic small clusters, *The journal of physical chemistry. A*, 2011, **115**, 1516–1520.

78  P. J. Hsu and S. K. Lai, Structures of bimetallic clusters, *The Journal of chemical physics*, 2006, **124**, 44711.

79  N. T. Wilson and R. L. Johnston, A theoretical study of atom ordering in copper–gold nanoalloy clusters, *J. Mater. Chem.*, 2002, **12**, 2913–2922.

80  B. Huang, H. Kobayashi, T. Yamamoto, S. Matsumura, Y. Nishida, K. Sato, K. Nagaoka, M. Haneda, S. Kawaguchi, Y. Kubota and H. Kitagawa, Coreduction methodology for immiscible alloys of CuRu solid-solution nanoparticles with high thermal stability and versatile exhaust purification ability, *Chem. Sci.*, 2020, **11**, 11413–11418.

81  P. P. Arquillière, I. S. Helgadottir, C. C. Santini, P.-H. Haumesser, M. Aouine, L. Massin and J.-L. Rousset, Bimetallic Ru–Cu Nanoparticles Synthesized in Ionic Liquids: Kinetically Controlled Size and Structure, *Top Catal*, 2013, **56**, 1192–1198.